\documentclass[%
 reprint,
 amsmath,amssymb,
 aps,
]{revtex4-1} 

\usepackage{graphicx}% Include figure files
\usepackage{dcolumn}% Align table columns on decimal point
\usepackage{bm}% textbf math
\usepackage{comment}

\usepackage[mathlines]{lineno}% Enable numbering of text and display math
\usepackage{multirow}
\usepackage{array}

\begin{document}

\preprint{APS/123-QED}

\title{Invisibility and perfect absorption of all-dielectric metasurfaces \\ originated from the transverse Kerker effect}

\author{Hadi~K.~Shamkhi$^1$}
\author{Andrey~Sayanskiy$^1$}
\author{Adri\`a Can\'os Valero$^1$}
\author{Anton~S.~Kupriianov$^{2}$}  
 \author{Polina~Kapitanova$^1$}
 \author{Yuri S. Kivshar$^3$}
 \author{Alexander~S.~Shalin$^1$}
\author{Vladimir~R.~Tuz$^{4,5}$}
\affiliation{$^1$ITMO University, St.~Petersburg 197101, Russia}
\affiliation{$^2$College of Physics, Jilin University, Changchun 130012, China} 
\affiliation{$^3$Nonlinear Physics Centre, Australian National University, Canberra ACT 2601, Australia}
\affiliation{$^4$International Center of Future Science, State Key Laboratory of Integrated Optoelectronics, College of Electronic Science and Engineering, Jilin University, Changchun 130012, China}
\affiliation{$^5$Institute of Radio Astronomy, National Academy of Sciences of Ukraine, Kharkiv 61002, Ukraine}

\date{\today}

\begin{abstract}
Dielectric metasurfaces perform unique photonics effects and serve as the engine of nowadays light-matter technologies. Here, we suggest theoretically and demonstrate experimentally the realization of a high transparency effect in a novel type of all-dielectric metasurface, where each constituting meta-atom of the lattice presents the so-called transverse Kerker effect. In contrast to Huygens' metasurfaces, both phase and amplitude of the incoming wave remain unperturbed at the resonant frequency and, consequently, our metasurface totally operates in the invisibility regime. We prove experimentally, for the microwave frequency range, that both phase and amplitude of the transmitted wave from the metasurface remain almost unaffected. Finally, we demonstrate both numerically and experimentally and explain theoretically in detail a novel mechanism to achieve perfect absorption of the incident light enabled by the resonant response of the dielectric metasurfaces placed in the vicinity of a conducting substrate. In the subdiffractive limit, we show the aforementioned effects are mainly determined by the optical response of the constituting meta-atoms rather than the collective lattice contributions. With the spectrum scalability, our findings can be incorporated in engineering devices for energy harvesting, nonlinear phenomena and filters applications.
\end{abstract}


\maketitle
%\tableofcontents
\section{\label{intr}Introduction}
The scattering of light by particles is one of the classical problems in the theory of electromagnetism and optics (see classical textbooks \cite{Kerker_book_1969, Hulst_book_1981, Bohren_book_1998} and the recent review \cite{Frezza_JOSAB_2018}). Initially, the scattering theory (known as Mie scattering theory) was developed for homogeneous spherical particles, for which exact analytical solutions can be derived in the form of multipole expansions. The solutions are obtained as an infinite series of partial spherical waves, called multipoles, where the contribution of each multipole to scattering is given by a numerical weight factor, called multipole moment. Each multipole moment is associated with a specific charge-current distribution. Furthermore, this exact solution has been extensively studied and successfully applied to describe scattering by particles of various shapes and compositions (see the comprehensive overviews of available theories in \cite{Mishchenko_book_2000, Kahnert_JQSRT_2003, Wriedt_inbook_2012} and references therein). Besides, considerable effort has been devoted into solving scattering problems with simpler and more compact mathematical formulations targeted for various practical applications \cite{Tretyakov_book_2017, Evlyukhin_PhysRevB_2016, PhysRevB.99.045424, Gurvitz2019, Alaee_AdvOptMat_2019}. 

From the standpoint of practical applications, it is often desirable to suppress scattering from particles in a given direction. In particular, the suppression of forward and backward scattering are of special interest. Typically, scattering suppression is achieved by realizing multilayered designs of particles \cite{Naraghi_OptLett_2015, Diaz_Avino_OE_2016} or covering them with special resistive (thin film) coatings. Such coatings can be made of lossy dielectrics \cite{Yamane_IEEETransCompat_2000} or metals \cite{Sureau_AntennaPropag_1967, Kildal_AntennaPropag_1996} depending on the wavelength of the scattered radiation. Recently, graphene has also been proposed as a coating material \cite{Riso_JOpt_2015, Naserpour_SR_2017, Fesenko_JOSAA_2018, Shcherbinin_JOSAB_2018}. While this approach is very effective, its implementation at the nanoscale remains a complex task with the current technological resources.

Fortunately, for uncoated particles, the scattering suppression can be also obtained. It appears naturally in magnetic particles, when particular conditions between the permittivity $\varepsilon$ and permeability $\mu$ of the particles are satisfied \cite{Kerker_JOSA_1983, Garcia-Camara_OptLett_2011}. For a small (\textit{subwavelength}) magnetic sphere, when the condition $\mu=\varepsilon$ holds, the backward scattering caused by the sphere appears eliminated, whereas the forward scattering is suppressed when the condition $\mu=-\varepsilon$ holds. This scattering suppression is known as Kerker effect, and from the viewpoint of the Mie scattering theory, it arises from the resonant overlapping electric dipole (ED) and magnetic dipole (MD) moments with the same magnitude (in particular, there is a near-zero backward scattering when ED and MD moments are in phase, and a strong backward scattering when they are out of phase). 

Although these conditions were obtained for a small magnetic sphere, similar scattering characteristics can be achieved for non-magnetic particles with large enough values of permittivity. For such particles the MD moment also makes a contribution to the scattering in the small-particle limit \cite{Kerker_book_1969, Kerker_JOSA_1983}, and thus, it can overlap the ED moment with the same magnitude, when the sphere radius is consciously adjusted. Moreover, there is a possibility for similar overlapping other higher-order moments (e.g. electric quadrupole (EQ) and magnetic quadrupole (MQ) moments and so on). The manifestation of interplay between the multipolar moments in the scattering by particles of an arbitrary shape under different radiation conditions is now referred to as \textit{generalized} Kerker effect \cite{liu2017multipolar,Terekhov_broadband} (see also extensions specific for the generalized Kerker effect listed in \cite{Liu_OptExpress_2018}). 

While the mechanism of suppression of forward and backward scattering on a single dielectric particle mediated by Kerker effect is well described, there is a special interest in arrangement of such particles into a lattice in order to design functional metasurfaces with desired reflection and transmission characteristics \cite{PhysRevB.99.045424}. In particular, one can utilize the generalized Kerker effect in the metasurface composed of dielectric particles to suppress reflection from the high-index substrate, and in this case, the metasurface serves as an antireflective coating \cite{baryshnikova2016plasmonic, Babicheva_JOSAB_2017}. One important class of such structures are reflectionless all-dielectric Huygens' metasurfaces operating on spectrally overlapped MD and ED moments oriented perpendicular to each other with equal magnitude \cite{Decker_AdvOptMat_2015, Babicheva2017}.

One significant advantage of all-dielectric metasurfaces is the possibility to fabricate them on a traditional Silicon-on-insulator (SOI) platform. Unfortunately, considering permittivity of the bare silicon in the visible part of spectrum desired for operation (with the typical diameter of particles in the range of $50-200$~nm), the Kerker effect conditions cannot be satisfied for spherical particles. Nevertheless, by adjusting diameter and thickness of cylindrical particles (disks) or width and height of rectangular particles (cubes), it is possible to bring corresponding moments into overlap. Moreover, metasurfaces based on silicon particles made in the form of disks or cubes are easier to fabricate at the nanoscale by utilizing modern lithography technologies.    

Although the technologies of manufacturing of silicon nanoparticles are well established today, production of all-dielectric metasurfaces at the nanoscale still requires expensive equipment and materials \cite{Yang_PhysRep_2017}. It can be also time-consuming. In this regard, correct modeling and prototyping of nanostructures become very important. Accordingly, in order to confirm developed concepts of all-dielectric nanostructures, quasi-optic (microwave) experiments can be used. For the microwave range, it is often much easier and cheaper to fabricate required samples with micron resolution using modern milling machines, laser cutting and 3D printing technologies. Moreover, the sources of the microwave radiation are also relatively inexpensive. 

In the present paper, we study and implement metasurfaces which allow the suppression of the reflected fields. In contrast to Huygens' metasurfaces, the transmitted light traverses the metasurface without perturbation of its amplitude and phase, therefore displaying lattice invisibility effect accompanied with strong non-trivial near-fields, that can also be interpreted as full transmission of light through the photonic structure. The constitutive building blocks (i.e. meta-atoms) of the invisible metasurfaces are ``transverse scatterers'' and have been extensively investigated in our previous work \cite{shamkhi_transverse}. We show here the duality of the Kerker and anti-Kerker conditions governing the transverse scattering of single scatterers still holds for invisible lattices in the sub-diffractive regime regardless of variations in the lattice parameter. For the  first time to our knowledge, we prove experimentally in the microwave range that both  amplitude and phase of the transmitted wave remain unaffected. The multipole decomposition semi-analytic approach of a single particle  scattering has been extended to decompose the reflection and transmission coefficients of a sub-diffractive lattice system in a homogeneous medium with or without the presence of a substrate. \par 
By exposing the coupling extent between multipole resonances, we find the zero-reflection point is not affected by variations in the lattice spacing within the sub-diffractive region. Finally, we present a different mechanism of reflection suppression from a dielectric metasurface placed near a conducting sheet. In this case, we show the light is fully absorbed at the magnetic quadrupole resonance by exploring the link between the resonant properties of the standalone dielectric cavity and the optical response of the metasurface.  

\section{\label{mod}Far-field properties of all-dielectric metasurfaces}
\label{sec:semianalytic}

\begin{figure}[t!]
\centering
\includegraphics[width=1.0\linewidth]{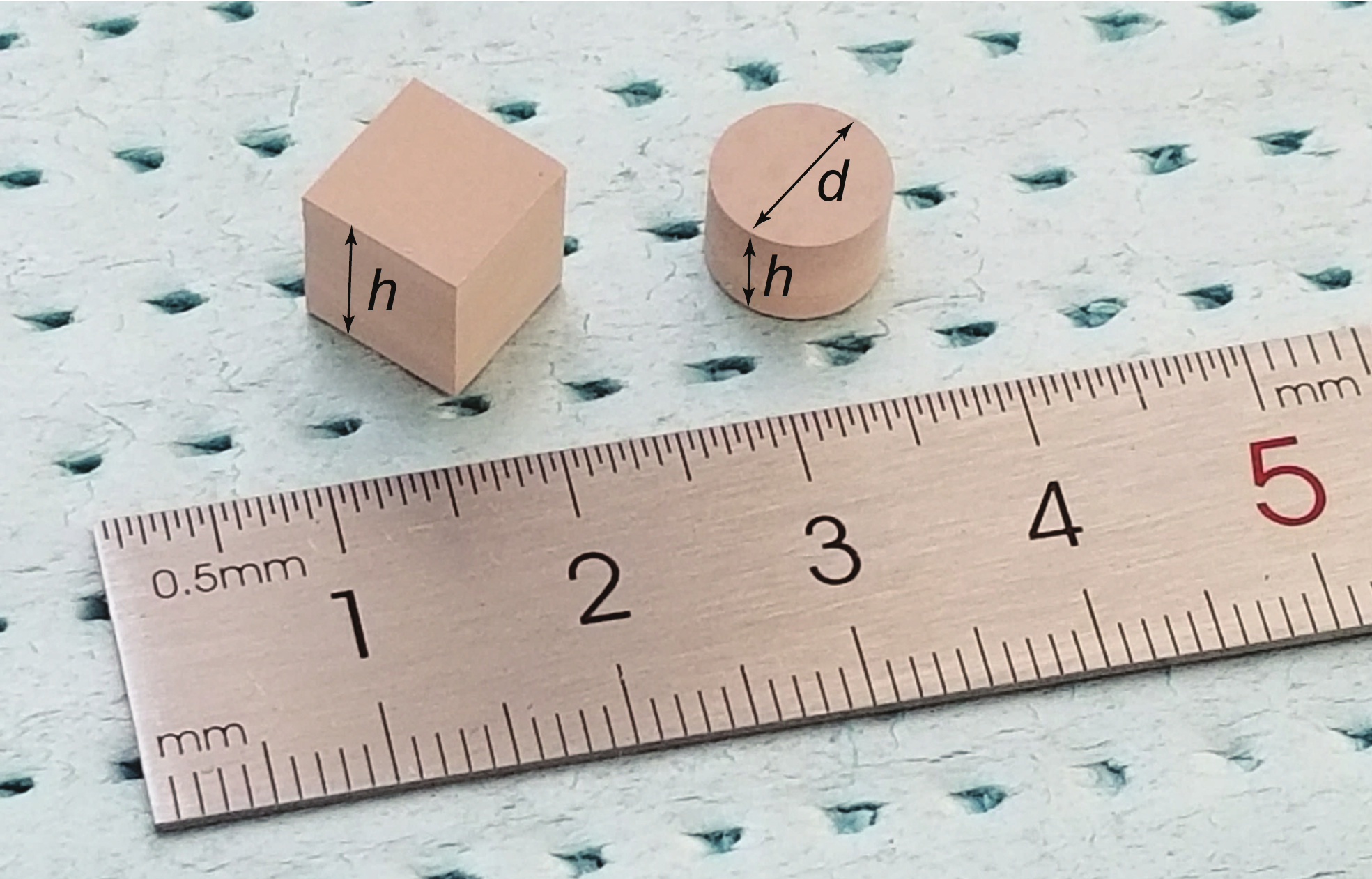}
\caption{Dielectric (ceramic) particles under study. The geometrical parameter $h$ represents the cubic particle edge and disk's height while $d$ gives disk's diameter.}
\label{fig:sketch}
\end{figure} 

Our metasurface configuration consists of periodically arranged dielectric particles in an infinite lattice in the  $x$-$y$ plane (Fig.~\ref{fig:sketch}). We assume the scattered field of each particle is fully characterized by the induced multipole moments up to quadrupoles. Thus, the total electric field outside the lattice is the summation of the scattering from all multipoles over the lattice geometry, 
\begin{equation}
\label{eq:totField}
\begin{split}
   & \textbf{E}(\textbf{r})\cong\textbf{E}_\textrm{inc}(\textbf{r}) \\&+\sum_{\textbf{r}_\parallel}^{\infty} \Big[\hat{g}^{Ep}\cdot\textbf{p}
    +\hat{g}^{Em}\cdot\textbf{m}+\hat{g}^{EQ}\colon\hat{Q}+\hat{g}^{EM}\colon\hat{M}\Big],
\end{split}
\end{equation}
characterized by the Green tensors \cite{Swiecicki2017}
\begin{equation}
\label{eq:GreenTensors}
\begin{split}
    &g_{ij}^{Ep}=\frac{1}{ \varepsilon_0}\partial_{ij} g_{0}(\textbf{r}-\textbf{r}_\parallel),\\
    &g_{ij}^{Em}=\frac{ik}{ c\varepsilon_0 }\epsilon_{isj}\partial_{s}g_{0}(\textbf{r}-\textbf{r}_\parallel),\\
    &g_{ijl}^{EQ}=-\frac{1}{ \varepsilon_0 }(\partial_{l}\partial_{ij}+\partial_{j}\partial_{il})g_{0}(\textbf{r}-\textbf{r}_\parallel),\\
    &g_{ijl}^{EM}=-\frac{ik}{ c \varepsilon_0 }(\epsilon_{ijm}\partial_{ml}+\epsilon_{ilm}\partial_{mj})g_{0}(\textbf{r}-\textbf{r}_\parallel),
\end{split}
\end{equation}
where $g_{0}(\textbf{r})$ is the scalar Green function, 
 $\textbf{r}$ denotes a point in space relative to the coordinate origin, $\textbf{E}_\textrm{inc}$ is the incident field, $ \epsilon_{isj}$ is the Levi-Civita symbol and $\partial_{ij}=\partial_{i}\partial_{j}-k^2\delta_{ij}$. The vector $\textbf{r}_\parallel = n\textbf{a}_x + m\textbf{a}_y $ represents the positions of the particles, where $\textbf{a}_x$, $\textbf{a}_y$ are lattice vectors, $n$, $m$ are integers. 
\begin{figure}[!t]
\centering
\includegraphics[width=1.0\linewidth]{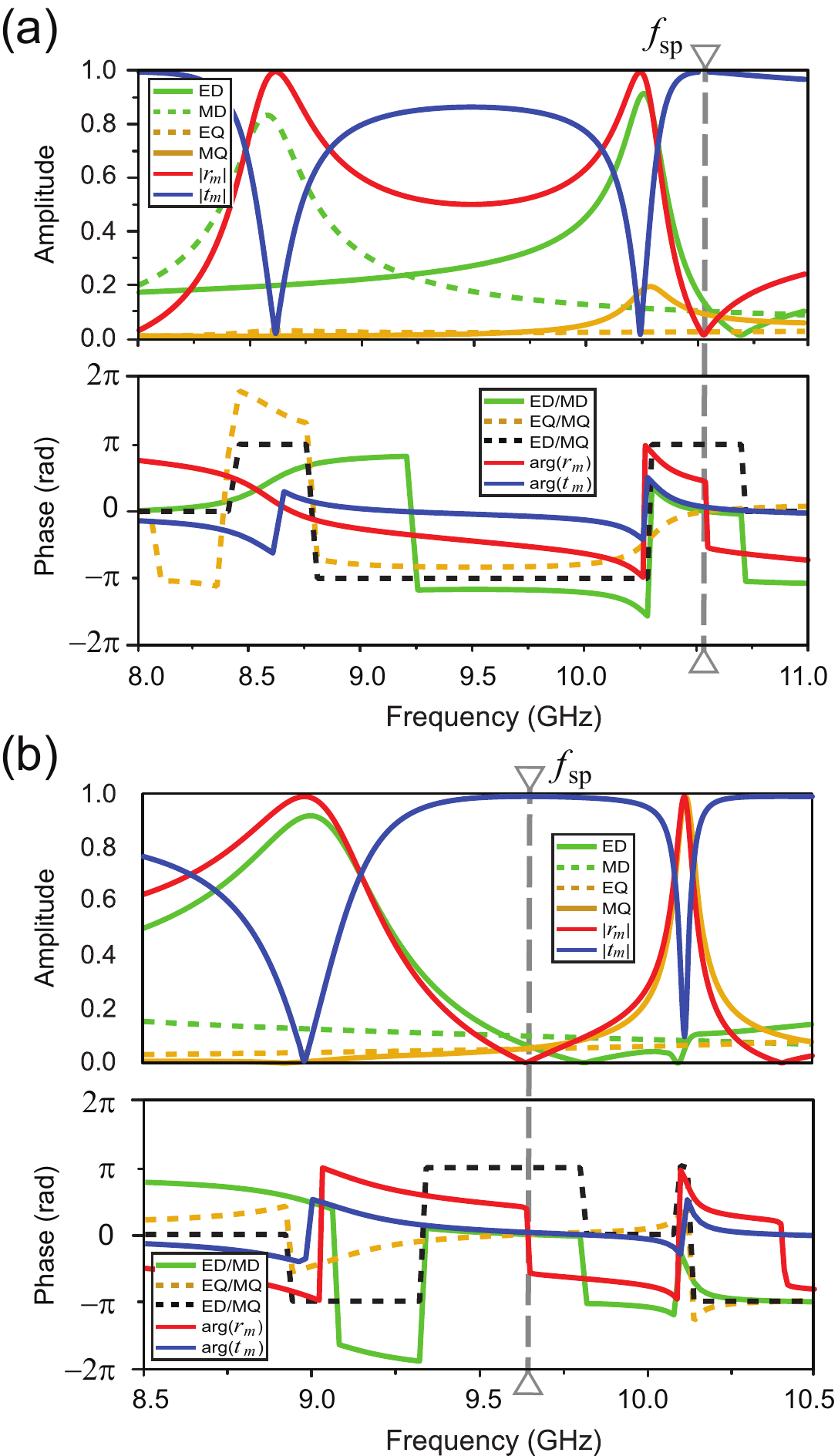}
\caption{Simulated reflection ($r_m$), transmission ($t_m$)  coefficients, and multipole decomposition of an all-dielectric lossless metasurface composed of (a) disk and (b) cubic particles. The decomposition is done via utilizing Eq.~(\ref{eq:vacuumR}) and the vertical lines indicate the spectral position of the transverse Kerker effect (invisibility effect). The geometrical and material parameters of the metasurfaces under study and their constituting building blocks are: (a) $d=8.0$~mm, $h=5.0$~mm, $a=20$~mm, $\varepsilon=23$; (b) $h=8.0$~mm, $a=22$~mm, $\varepsilon=21$.}
 %$; $a=22$~mm
\label{fig:simulated}
\end{figure}
The term $\hat{g}^{Ep}\cdot\textbf{p}$ gives the sum of the field scattered by the EDs. Similarly, $\hat{g}^{Em}\cdot\textbf{m}$, $\hat{g}^{EQ}:\hat{Q}$ and $\hat{g}^{EM}\colon\hat{M}$ represent the contribution of the MD, EQ and the MQ to the total field, respectively. The summation in Eq.~(\ref{eq:totField}) includes evanescent and propagating waves. The latter correspond to open diffraction channels contributing to the far field. In the sub-diffractive region, the reflection and transmission of the lattice are characterized by the first order diffraction, providing non-zero contributions to the far-field \cite{Garcia_modernPhysics, goodmanFourier}.
Knowledge of the field outside the lattice allows the formulation of the non-Fresnel reflection $z > 0$ and transmission $z <0 $ coefficients of the lattice as 
\begin{equation}
\label{eq:latticeref}
   r_{m}=\frac{(\textbf{E}-\textbf{E}_\textrm{inc})\cdot \textbf{E}_\textrm{inc}^{*} }{\textbf{E}_\textrm{inc}\cdot \textbf{E}_\textrm{inc}^{*}},~
    t_{m}=\frac{\textbf{E}\cdot \textbf{E}_\textrm{inc}^{*} }{\textbf{E}_\textrm{inc}\cdot \textbf{E}_\textrm{inc}^{*}}.
\end{equation}
These coefficients are complex numbers even for lossless dielectric particles and contain the characteristics of the lattice periodicity. Throughout this work, we consider the metasurfaces under study are illuminated by a normally incident, linearly polarized plane wave $\mathbf{E}_\textrm{inc}=E_{0}e^{ikz}\mathbf{x}$, where $\textbf{x}$ is a unit vector directed along the $x$ axis. Eq.~(\ref{eq:latticeref}) can then be reduced to \cite{PhysRevB.99.045424} 
\begin{equation}
\label{eq:vacuumR}
\begin{split}
   & r_{m}\cong \frac{ik}{2E_0 A \varepsilon_0 }  \Big(p_{x}+\frac{1}{c}m_{y}-\frac{ik}{6}Q_{xz}-\frac{ik}{2c}M_{yz}\Big),
   \\& t_{m}\cong1+ \frac{ik}{2E_0 A \varepsilon_0 } \Big(p_{x}-\frac{1}{c}m_{y}+\frac{ik}{6}Q_{xz}-\frac{ik}{2c}M_{yz}\Big),
\end{split}
\end{equation}
where $c$ is the speed of light in vacuum and $A=\vert\textbf{a}_{x}\times\textbf{a}_{y}\vert$ is the area of the unit cell. The last equations link the scattering pattern of a single particle and the far-field wavefront of the lattice, thus allowing to study the characteristics of the field emitted by the lattice by means of the multipole decomposition of the field scattered by a single meta-atom of the lattice. We previously found the conditions at which the forward and backward scattering of a standalone scatterer are nearly suppressed simultaneously~\cite{shamkhi_transverse}. Similar conditions can be obtained for a fully transparent lattice in a homogeneous medium directly by setting the lattice reflection and its contribution to transmission to zero, $r_{m}=0$ and $1-t_{m}=0$, resulting in
\begin{equation}
\label{eq:conditions}
p_{x}=\frac{m_{y}}{c},~~ \frac{Q_{xz}}{3}=\frac{M_{yz}}{c},~~p_{x}=-\frac{kM_{yz}}{2ic}.
\end{equation}
In the previous, we have assumed the scattering is fully characterized by the multipole contributions up to the MQ, however these conditions can be further extended to include higher order multipoles. The first term of Eq.~(\ref{eq:conditions}) is the well-known Kerker condition for dipoles and the second term is the Kerker-like condition for quadrupoles. The last term is of particular interest, since it suggests that the coherent dipoles are in a $\pi$ phase relation with respect to the coherent quadrupoles or the anti-Kerker condition of the dipole-quadrupolar scatterers. Simultaneous fulfillment of the Kerker and anti-Kerker conditions leads to the redirection of the scattered power to the lateral directions and enhanced suppression of the contributions to reflection and transmission \cite{shamkhi_transverse} leaving the incident wave almost unperturbed.

We have further validated our analytical predictions against numerical simulations and experiments. In accordance with our available experimental means, we have chosen the microwave range of the spectrum ($8-15$ GHz) to demonstrate the manifestation of the invisibility effect. We consider particle arrangements constituted of a low-loss, high-permittivity microwave ceramic, manufactured in the shapes of disks and cubes (Fig.~\ref{fig:sketch}). 

For the metasurfaces constituted of such particles we have performed numerical simulations in COMSOL Multiphysics\textsuperscript{\textregistered}. The electromagnetic response of the metasurfaces was obtained by imposing Floquet-periodic boundary conditions on four sides of the unit cell to simulate an infinite two-dimensional array of particles. Figure~\ref{fig:simulated} shows the amplitudes and phases of the simulated reflection and transmission coefficients for two particular metasurfaces constituted of disks [Fig.~\ref{fig:simulated}(a)], and cubes [Fig.~\ref{fig:simulated}(b)], organized in a square lattice with spacing $a=\vert \textbf{a}_{x}\vert=\vert \textbf{a}_{y}\vert$. The simulations confirm that in the chosen frequency band areas of complete transparency indeed exist. We mark corresponding resonant frequencies by vertical dashed lines (they are $f_\textrm{sp}=10.5$ GHz and $f_\textrm{sp}=9.67$ GHz for the array of disks and the array of cubes, respectively).

We decompose the reflection spectra in Fig.~\ref{fig:simulated}. As a first remark, we note that the near-zero reflection point is located in the vicinity of the Fano-like profile of the ED resonance. Secondly, the MQ resonance appears on the red side of the near-zero reflection point for the array of disks and on the blue side for the cubic array, while the other multipoles are nonresonant. Consequently, conditions (\ref{eq:conditions}) are nearly satisfied; the four leading multipoles have comparable amplitudes at the zero reflection point. The dipoles are in-phase (Kerker) and the same situation takes place for quadrupoles (generalized Kerker). The last term of conditions (\ref{eq:conditions}) as depicted by the black dashed line shows the coherent dipoles are in anti-phase with the coherent quadrupoles. In contraposition to the well-known Huygens' metasurfaces where strong forward scattering results in a phase difference between the incident and transmitted fields, here we notice both waves are in phase (blue lines), meaning this kind of metasurfaces are extraordinary transparent with almost no change of amplitude or phase of the transmitted wave.

\section{Effect of lattice spacing and material losses}

Variations in the lattice spacing influence the optical response of a metasurface in an equivalent way to changes in the shape and composition of the meta-atom. The spectral position of the electric and magnetic modes in a lattice can be tailored by adjusting spacing in one of the lattice axes while keeping the other constant, in such a fashion that one can bring both resonances to overlap and suppress reflection \cite{Staude2013, Babicheva2017}. We show here that although the Fano profile of the ED and MQ resonances are spectrally shifted in different directions, the frequency of the suppressed reflection corresponding to their shared dip point is fixed on the spectrum. We restrict ourselves to the consideration of the array of cubes, however the discussion could be applied to all similar scenarios that satisfy the conditions (\ref{eq:conditions}). 

We impose the constraints that the lattice is axially symmetric for normally incident light.  The effective field acting on each particle centre is the summation of the incident field and all of the multipoles radiations in the lattice. Furthermore from the definition of the Green tensors that describe multipole scattering [Eq.~(\ref{eq:GreenTensors})], the coupling between the ED and both MD and EQ are cancelled out due to the lattice symmetry. The MQ on the other hand is very small in the region where the ED Fano resonance appears, hence we count only for the ED-ED coupling which can be formulated over lattice geometry as 
\begin{equation}
    \label{eq:1dgeneral}
    \begin{split}
    %=\sum_{n,m} \hat{g}^{Ep}(n a_x,m a_y,0) \\&
         \hat{d} =\sum_{n\neq 0} \hat{g}^{Ep}(n \mathbf{a}_x,0,0)+\sum_{n}\sum_{m\neq 0} \hat{g}^{Ep}(n \mathbf{a}_x,m \mathbf{a}_y,0),
    \end{split}
  \end{equation}
where $\hat{d}$ is a second order tensor representing the dipole-dipole coupling. Here the summation is split over the lattice geometry to a summation over one of the coordinate axis (the first part) and a summation over all the remaining lines in the lattice. The excitation wave setup restricts the excited electric multipole components to the plane of incidence \cite{Evlyukhin_PhysRevB_2016}. In the case of $x$-polarized incident field, only the $p_x$ component of the ED is induced. Also, due to the lattice symmetry, all components of the periodic tensor $\hat{d}$ except the diagonals are zero. We write \cite{Evlyukhin2010,Markel_2005, Liu_PolarizationIndependent, Swiecicki2017}
\begin{equation}
\label{eq:effectiveMoment}
    p_{x}\cong \frac{\alpha_{p} E_{0}}{1-\alpha_{p} d_{xx}},
\end{equation}
with
\begin{equation}
    \label{eq:1d}
    \begin{split}
         {d}_{xx}&\cong-\frac{2 k^{2}}{a}\Big(\ln[2-2\cos(ka)]+i\pi-ka\Big)\\&+\sum_{n\neq 0}\frac{2inka- 1}{n^{3}a^{3}} e^{inka}+\sum_{n}\sum_{m\neq 0}\frac{\omega_{n}^{2}}{2 a^{2}} K_{0}(-i m a \omega_{n}),
    \end{split}
\end{equation}
where $\alpha_{p}$ is the ED polarizability of the individual particle, $K_0$ is the Bessel function of the second kind, and $\omega_{n}=\sqrt{k^{2}-(2\pi n/a)^{2}}$. The lattice effect (coupling) is determined by the spacing and rises rapidly closer to and around the first diffraction order \cite{Markel_2005, Liu_PolarizationIndependent}. This spectral position where the wavelength equals the lattice spacing $\lambda=a$ is referred to as the first diffraction order or Rayleigh anomaly. The energy confinement in the first diffraction order causes the coupling to diverge instantaneously. Upon inspection of Eq.~(\ref{eq:effectiveMoment}), one can see the effective multipole moments change when the product ($\alpha_{p} d_{xx}$) in the denominator becomes significant. Hence both the ED polarizability and the coupling are equally relevant and need to be considered in order to determine the limitations of the fixed-position reflection. 

\begin{figure}[t!]
\centering
\includegraphics[width=1\linewidth]{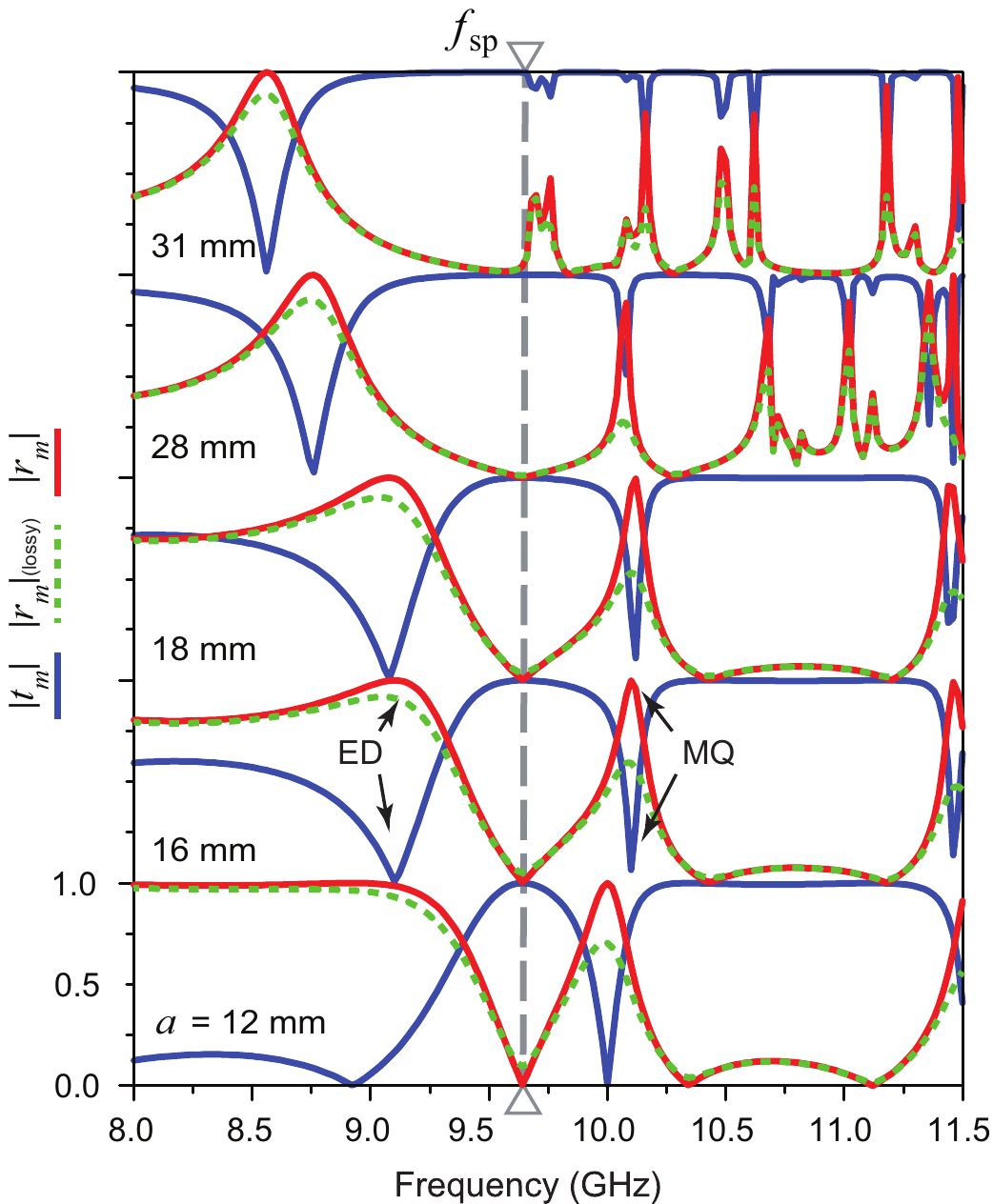}
\caption{Calculated reflection and transmission coefficient spectra for different lattice spacings of a square lattice constituted of our ceramic cubes. Dashed green curves show the reflection from the metasurface where the real material losses have been taken into account ($\tan\delta=8 \times 10^{-3}$). The vertical grey dashed line indicates the fixed spectral position of the reflection suppression point between the electric dipole (ED) and the magnetic quadrupole (MQ) resonances. The geometrical parameters of the particles are the same as in Fig.~\ref{fig:simulated}.}
\label{fig:symmetric}
\end{figure}

Figure~\ref{fig:symmetric} shows the reflection and transmission spectra for different lattice spacings, where we have considered the array of cubes with and without accounting for material losses. In the sub-diffractive region, the low reflection point at the transverse scattering ($f_\textrm{sp} = 9.67$~GHz, $\lambda_\textrm{sp} = 31$~mm) is fixed while other low reflection points are shifted. Also there are opposite spectral displacements of the ED and MQ modes leading to consequent broadening or narrowing of the low reflection region. It was previously reported \cite{Markel_2005} that the evolution of ED-ED coupling along 1D arrays results in an increase of the interaction between dipoles with a wider spacing, while the diffraction orders work as a divergence and reset the points. The same goes for 2D lattices [see the last term in Eq.~(\ref{eq:1d})], since it can be considered as an infinite summation of 1D lattices \cite{Simoviski_appliedPhysics}. We notice here that the Fano profile of ED is deformed and weakened as the ED-ED coupling increases. The spectral regions with a resonant particle polarizability enables a dramatic change in the reflection behavior since the term ($\alpha_{p} d_{xx}$) becomes significant and influences the effective dipole moment. Therefore, as can be seen in Fig. \ref{fig:symmetric}, the speed of shifting is not homogeneous. Nevertheless, since the zero reflection point has a very low polarizability and, therefore, is fixed within the sub-diffractive range of spacing.  Note, if spacing is very small compared to the wavelength $a \ll \lambda$, the higher order multipoles are excited and the aforementioned discussion is no longer applicable.

Now let us examine the lattice response beyond the first diffraction order ($\lambda=31$~mm). In this case there is a divergent coupling along with the Fano profile dip. If spacing is beyond this limit, a new lattice forced ED Fano profile leads to revoking the zero-reflection point \cite{Liu_PolarizationIndependent}. Noteworthy for oblique incidence, the first diffraction order is effectively moved to shorter wavelengths \cite{RevModPhys.82.2257}.

Contrary to the ED, the MQ resonance shifts to the blue side of the spectrum when lattice spacing is increased (Fig.~\ref{fig:symmetric}).  However, according to the characteristics of the periodic Green's tensors, the ED-MQ (MD-EQ) coupling is non vanishing due to the lattice symmetry \cite{Swiecicki2017}. For instance, the coupling between MD and EQ in a symmetric lattice of finite size in a homogeneous medium causes a shift in the EQ resonance \cite{Babicheva_ACS}. In the current profile, the ED contribution to the reflection and transmission at the frequency of the MQ resonance is sufficient   (Fig.~\ref{fig:simulated}) and their coupling causes a blue shift in the MQ resonance. The opposite displacements of the ED and MQ resonances lead to a broadening of the low-reflection area around the zero reflection point.

We now proceed to analyze the effect of the particle inherent losses on the reflection suppression. From Fig.~\ref{fig:symmetric} (see the dashed green line), we find the reflection to become slightly enhanced at the point of interest, however no change in its spectral position can be appreciated. In fact, losses decrease the $Q$ factor of the Fano resonance. Resonances in subwavelength open dielectric resonators are governed by the imaginary part of the permittivity~\cite{Silveirinha2014,Rybin2017} as $Q \backsim \Re (\varepsilon)/\Im (2\varepsilon)$. With the current permittivity of ceramic particles $\varepsilon=21+0.168i$ and $Q\backsim 63$, we can conclude that, for example, if our lattice was constituted of silicon particles in the near infrared region, where the quality factor can exceed several thousands, our spectra would display an almost zero reflection point.
\section{\label{rep}Experimental demonstrations}

In order to verify the theoretical and numerical predictions, we manufactured and tested two square metasurfaces, each one containing $18 \times 18$ unit cells and having the side length of 400~mm. Their unit cells correspond to those presented in Figs.~\ref{fig:simulated}(a) and \ref{fig:simulated}(b) for the metasurfaces composed from disks and cubes, respectively. As a dielectric material the Taizhou Wangling TP-series microwave ceramic characterized by the relative permittivity $\varepsilon= 23$ for disks and $\varepsilon= 21$ for cubes has been used. The loss tangent for the ceramic is $\tan\delta \approx 6 \times 10^{-3}$ at 10~GHz. The dielectric particles with the sizes mentioned in the caption of Fig.~\ref{fig:simulated} were fabricated with the use of precise mechanical cutting techniques. To arrange them into a lattice, an array of holes was milled in a custom holder made of a Styrofoam material whose permittivity is $\varepsilon_s= 1.05$ and thickness of the plate is $h_s = 20.0$~mm. Further, its complex reflection and transmission coefficients were measured in the same frequency range as in the simulations using the common technique when the measurements are performed in the radiating near-field region and then transformed to the far-field zone. 

The measurements procedure as well as experimental setup are described in details in Refs.~\onlinecite{Sayanskiy2019PRB} and \onlinecite{xu_AdvOptMat_2019}. During the investigation the prototype was fixed on the $2.0$~m distance from a rectangular linearly polarized broadband horn antenna. The antenna generated a quasi-plane-wave with required polarization. The antenna was connected to the first port of the Keysight E5071C Vector Network Analyzer (VNA) by a 50~Ohm coaxial cable. To detect the electric field, an electrically small dipole probe connected to the second port of the VNA was used. 

\begin{figure}[!t]
\centering
\includegraphics[width=1.0\linewidth]{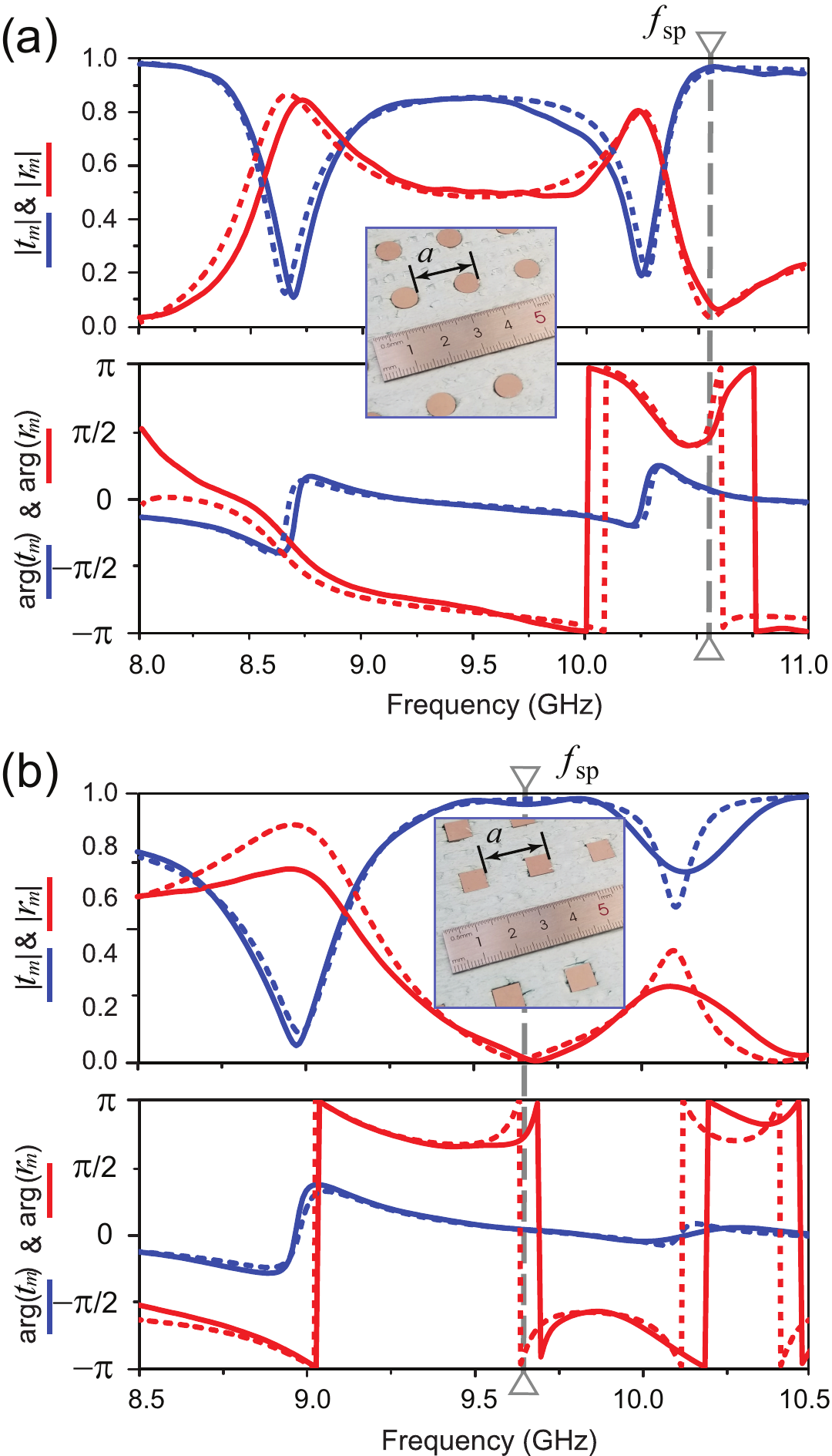}
\caption{Simulated (dashed lines) and measured (solid lines) transmission ($t_m$) and reflection ($r_m$) coefficients of the all-dielectric metasurface composed of (a) disk and (b) cubic particles. The insets demonstrate fragments of the metasurface prototypes. In the simulation the substrate is modeled as a lossless dielectric with near unity refractive index, while actual material losses in ceramic particles are taken into account, where (a) $\varepsilon = 23 + i0.138$, (b) $\varepsilon = 21 + i0.168$, and all geometrical parameters of the particles and lattice spacing are the same as in Fig.~\ref{fig:simulated}.}
\label{fig:measured}
\end{figure}

The measured transmission and reflection coefficients for both metasurfaces are collated in Fig.~\ref{fig:measured}. For comparison, additional numerical simulations were performed taking into account the measured deviations in metasurface dimensions from the nominal values appeared in fabrication as well as losses in the particles. The presented data demonstrates very good agreement between measurements and simulations, where one can see discussed above areas of the complete transparency.

\section{Metasurfaces placed on a~conducting substrate}

For various practical realizations, where the lattice is to be placed on a substrate, we extend the semi-analytic derivation of the reflection and transmission decomposition of the lattice shown in Sec.~\ref{sec:semianalytic} to 
account for a substrate. We found the transverse scatterers gradually transform to Kerker scatterers on high-index  substrates \cite{shamkhi_transverse}, and the broadband Huygens' metasurfaces can be realized as a result of this mechanism \cite{Decker_AdvOptMat_2015}. Nonetheless, in this section we discuss the transformation of the considered metasurface to a fully absorbing lattice with the addition of a conducting substrate (a PEC sheet in the microwave range). Analytic results will be supported by numerical calculations and experimental measurements. 
\begin{figure}[t]
\centering
\includegraphics[width=1.0\linewidth]{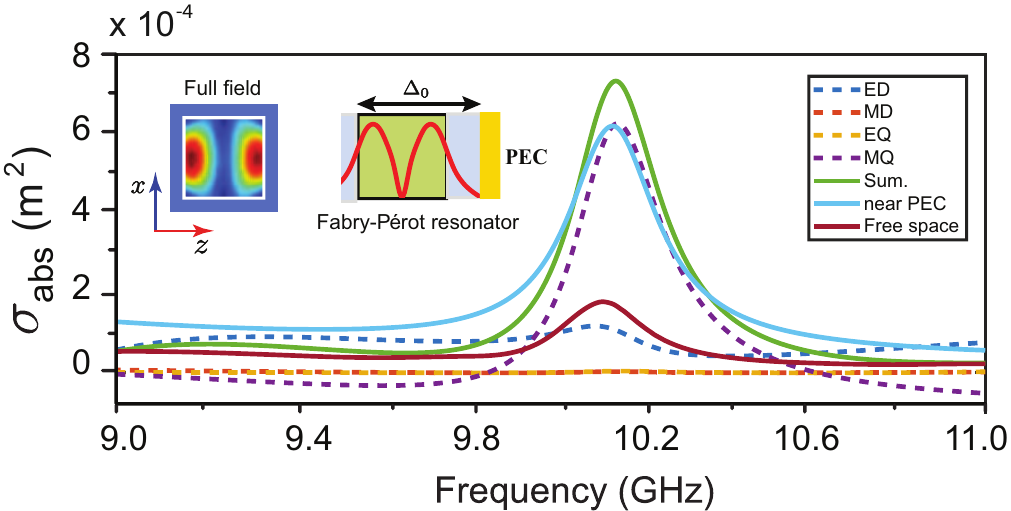}
\caption{Absorption cross section and its approximated multipole decomposition of the standalone ceramic cube in the two situations presented in the text, i.e. free space and placed 3~mm above the PEC sheet. The left inset shows the calculated electric field distribution associated with the MQ at resonance for the cube above the PEC sheet (the point of maximal absorption), and the right inset depicts the electric field amplitude of the $m=5/2$ eigenmode for the equivalent 1D Fabry-P\'erot resonator. The material and geometrical parameters of the cube are the same as in the metasurface in Fig.~\ref{fig:measured}(b).}
\label{fig:absorption}
\end{figure}
We start our discussion with an investigation of the spectral behavior of an isolated meta-atom, i.e. a single cube (introduced in Fig.~\ref{fig:sketch}) deposited over a PEC substrate. For simplicity of experimental part and numerical calculations hereinafter we add 3 mm gap (much smaller than the wavelength) between the cubes and the substrate.  In order to provide a better physical intuition, let us follow Ref.~\onlinecite{Lalanne2018} and consider only spatial variations along the height of the cavity. In the 1D case, an open cavity can be seen as a lossy dielectric Fabry-P\'erot resonator of refractive index $n=\sqrt{\varepsilon}$. After excitation with an external pulse (the incident wave), the resonator supports a series of standing wave-like field patterns whose excitation depends on the number of effective wavelengths that can be introduced inside the cavity of length $\Delta_0$ (see inset of Fig.~\ref{fig:absorption}). In the general case, the walls of the resonator have non-zero transmittance, and thus allow the confined modes to leak to the environment in the form of outgoing plane waves. When the internal field interferes constructively with itself, mode resonances occur. The latter take place only when the phase shift induced in the circulating wave after reaching the second wall and returning to the first accounts for $2\pi$. Using this fact, it is straightforward to show that the resonant frequencies for a lossless dielectric Fabry-P\'erot resonator are \cite{Ismail2016}
\begin{equation}
f_m=\frac{cm}{2n\Delta_0},
\label{eqn:eigenfrequencies}
\end{equation}
where $m$ is an integer or half-integer number accounting for the different resonance modes \cite{BornWolf_1999}. If losses are considered, an imaginary term needs to be added to the expression for $f_m$ \cite{Lalanne2018}, but the real part remains the same as in Eq.~(\ref{eqn:eigenfrequencies}). Higher values of $m$ allow to contain additional quarter wavelengths inside the resonator. For example, for the first three $m=1/2,1,3/2$, a quarter wavelength, half-wavelength and three quarters of a wavelength can be fitted in the length $\Delta_0$, respectively. In the general case, the effective wavelengths that can be fitted in the resonator for a given $m$ are $\Delta_0/\lambda_\textrm{eff}=m/2$, where $\lambda_\textrm{eff}=\lambda/n$, $\lambda$ is the wavelength in free space. \par
In Fig. \ref{fig:absorption} we show the numerically calculated absorption cross section of the cubic particle above PEC substrate, its multipole decomposition, and insets showing the link between the aforementioned Fabry-P\'erot mode and the resonant MQ.

 The individual contribution of the leading multipoles was calculated approximately as the difference between the extinction and scattering cross sections \cite{Evlyukhin_PhysRevB_2016, Gurvitz2019}.
First, we considered the exact expressions for the extinction cross sections of the electric and magnetic dipoles and quadrupoles  \cite{Miroshnichenko_magnetoelectric,Evlyukhin_PhysRevB_2016}

\begin{equation}
\label{eq:ext}
\begin{split}
     &\sigma _\textrm{ext}^{p}=\frac{\omega }{2{{I}_{0}}}\operatorname{Im}\left\{ \mathbf{p}\cdot \mathbf{E}_\textrm{exc}^{*}({{\mathbf{r}}_{0}}) \right\},\\&
     \sigma _\textrm{ext}^{m}=\frac{\omega {{\mu }_{0}}}{2{{I}_{0}}}\operatorname{Im}\left\{ \mathbf{m}\cdot \mathbf{H}_\textrm{exc}^{*}({{\mathbf{r}}_{0}}) \right\},\\&
     \sigma _\textrm{ext}^{Q}=\frac{\omega }{12{{I}_{0}}}\operatorname{Re}\left\{ i\hat{Q}:{{\left. \nabla \mathbf{E}_\textrm{exc}^{*} \right|}_{\mathbf{r}={{\mathbf{r}}_{0}}}} \right\},\\&
     \sigma _\textrm{ext}^{M}=-\frac{1}{2{{I}_{0}}}\operatorname{Re}\left\{ {{\left. \left[ \nabla \times \left( \hat{M}\cdot \nabla  \right) \right]\cdot \mathbf{E}_\textrm{exc}^{*} \right|}_{\mathbf{r}={{\mathbf{r}}_{0}}}} \right\},
\end{split}
\end{equation}
where $\omega$ is the angular frequency, ${{I}_{0}}$ is the incident energy flux and $\mathbf{E}_\textrm{exc}^{*}({{\mathbf{r}}_{0}})$ and $\mathbf{H}_\textrm{exc}^{*}({{\mathbf{r}}_{0}})$ are the complex conjugates of the excitation electric and magnetic fields evaluated at the center of the particle ${{\mathbf{r}}_{0}}$.
The expressions above are valid for any excitation field and particles of arbitrary shape. For an $x$-polarized plane wave normally incident on a PEC substrate along the $z$ axis, the exciting field which needs to be substituted in Eq.~(\ref{eq:ext}) corresponds to the sum of the incident and reflected waves. 
Second, the individual contributions of each multipole to absorption are assumed to be:

\begin{equation}
\label{eq:opticTherom}
    \sigma _\textrm{abs}^{L}=\sigma _\textrm{ext}^{L}-\sigma _\textrm{sca}^{L},
\end{equation}
where $L$ denotes any of the leading multipole moments. The formulae for each term  in the scattering cross section decomposition is taken from Ref.~\onlinecite{Gurvitz2019}. Therefore, the approximate total absorption cross section can be written in the form,

\begin{equation}
\label{eq:absDecomp}
    {{\sigma }_\textrm{abs}}\cong\sigma _\textrm{abs}^{p}+\sigma _\textrm{abs}^{m}+\sigma _\textrm{abs}^{Q}+\sigma _\textrm{abs}^{M}.
\end{equation}
Here we neglect magnetoelectric coupling between the multipoles due to the PEC \cite{Miroshnichenko_magnetoelectric}, and other cross-terms between the multipoles. However, Fig.~\ref{fig:absorption} shows that Eq.~(\ref{eq:absDecomp}) is in good quantitative agreement with the numerically calculated absorption cross section obtained with COMSOL Multiphysics indicating the validity of the utilized approach in the considered spectral range.

Consequently, in Fig.~\ref{fig:absorption} we observe a strong enhancement of the absorption peak centered at the eigenfrequency in comparison with the free-space scenario. Moreover, the induced magnetic quadrupole responsible for this enhancement is fully driven by the eigenmode (standing wave) of the Fabry-P\'erot resonator. This is confirmed by the identical internal field distributions of the mode and the one obtained from numerical simulations (insets of Fig. 5). Thus, simple physics of 1D open resonators suffices to qualitatively describe the resonant behavior of the single nanoparticle on PEC, as well as the metasurface, as will be discussed in short. 
\begin{figure}[t]
\centering
\includegraphics[width=1.0\linewidth]{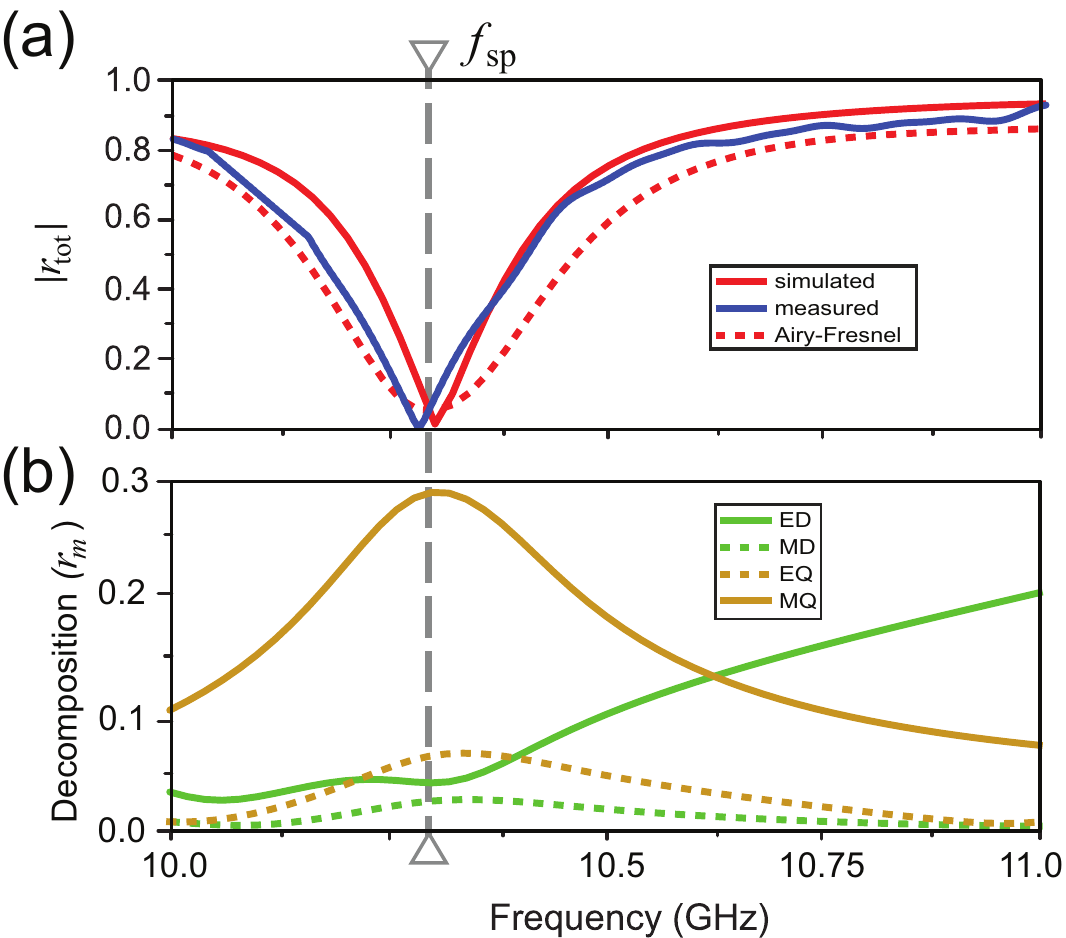}
\caption{(a) Simulated and measured total reflection coefficient of the all-dielectric metasurface ($r_\textrm{tot}$) composed of cubic particles and placed 3 mm above the PEC substrate. (b) The multipole decomposition of the first term of Eq.~(\ref{eq:reflectionPec}) ($r_m$) which represents the lattice direct contribution to the total reflection. Note here $r_{m}$ has been calculated taking into account the total excitation field acting on the particles. The material and geometrical parameters of the metasurface are the same as in Fig.~\ref{fig:measured}(b).}
\label{fig:screen}
\end{figure}

With the physical intuition gained by the investigation of the single meta-atom, we now turn our attention towards the case of the metasurface. Following the analytical treatment from Sec.~\ref{sec:semianalytic}, modelling the metasurface as a plane crossing the center of the particles with non-Fresnel reflection and transmission coefficients (\ref{eq:vacuumR}), and the method given at \cite{,Baranov_Applied,Shalin2010,Shalin_Spectroscopy} we can apply the well-known Airy-Fresnel formulae to study the reflection from the whole system. It is similar to the 1D Fabry-P\'erot model, but is suitable for metasurfaces since it takes into account the mutual particle interaction and provides useful information on each multipole contribution.

Now, $\Delta_h$ is the distance from the substrate interface to the plane, where point-like multipoles are localized (particle centers), considered as an imaginary interface \cite{Baranov_Applied, Shalin2010, Shalin_Spectroscopy}. These two boundaries now play the role of the walls of a resonator. The reflection and transmission coefficients of the whole system take the form: 
\begin{equation}
\label{eq:totreflection}
   r_\textrm{tot}=r_{m}+\frac{r_{s} t_{m}^{2} e^{2ik \Delta_h} }{1-r_{s} r_{m}~ e^{2ik \Delta_h}},
\end{equation}
and
 \begin{equation}
\label{eq:tottransmission}
     t_\textrm{tot}=\frac{t_{s} t_{m}}{1-r_{s} r_{m}~ e^{2ik \Delta_h}},
\end{equation}
where $r_\textrm{tot}$ and $t_\textrm{tot}$ are the overall system reflection and transmission with the substrate, and $r_{s}$ and $t_{s}$ are Fresnel coefficients of the substrate for the normally incident wave. When a PEC is utilized as the substrate (i.e the reflection of the lower wall of our Fabry-P\'erot resonator is $-1$), the total reflection is simplified to: 
\begin{equation}
\label{eq:reflectionPec}
    r_\textrm{tot}=r_{m}-\frac{t_{m}^{2} e^{2ik \Delta_h} }{1+r_{m} e^{2ik \Delta_h}},
\end{equation}
where $r_{s}=-1$ and $t_{s}=0$ have been substituted. Therefore, to cancel the reflection and absorb all incident light, the antireflection condition should be fulfilled, and direct reflection from the lattice and the substrate aided reflection (the second term of Eq. (\ref{eq:reflectionPec}) should be in a $\pi$  phase relation having equal amplitudes \cite{BornWolf_1999}. 

Let us utilize formula (\ref{eq:reflectionPec}) to study the reflection from the metasurface on a substrate. Figure \ref{fig:screen} shows the results of the analytical calculations together with the measured reflection and the results of simulations in COMSOL Multiphysics. Noteworthy, the results are very close to each other validating the suggested approach. 

The multipolar decomposition (\ref{eq:reflectionPec}) of the $r_{m}$ gives us the particular contributions of the multipoles to the reflection. As it is clearly seen from Fig. 6(b), the resonant MQ is once again responsible for the reflection dip (absorption peak, because PEC interface do now allow for non-zero transmission).  The spectral position of the MQ resonance at $f_\textrm{sp}=10.25$ GHz is slightly shifted in comparison to the free space case shown in Fig.~\ref{fig:simulated} which is at $f_\textrm{sp}=10.17$ GHz. However both positions are in the vicinity of the original MQ resonance of the standalone cubic particle (see Fig.~\ref{fig:absorption}), underlining that the resonant modes of an isolated meta-atom provide valuable insight into the optical response of the metasurface.

\section{\label{concl}Conclusions}

We have studied, both theoretically and experimentally, a novel class of all-dielectric Mie-resonant metasurfaces governed by the transverse Kerker effect. Such metasurfaces demonstrate a complete transparency similar to Huygens' metasurfaces, but they experience 
zero phase shift between the incident and transmitted waves. We have clarified the underlying physics of this effect, and we have formulated the specific conditions for which both reflection and transmission coefficients are suppressed simultaneously for square lattices composed of  Mie-resonant dielectric particles. This type of optical response occurs when the coherent dipole modes and coherent quadrupole modes excited in the dielectric particles satisfy the so-called generalized Kerker condition, and they possess a phase difference between each others. When these conditions are satisfied, the metasurface becomes absolutely transparent, and both amplitude and phase of a transmitted wave are completely unaffected. We have demonstrated this effect experimentally for microwave frequencies. Furthermore, the coupled multipoles in in the lattice do not affect the invisibility point. We have shown that a variation of the lattice spacing within the subdiffractive scattering regime results only in either narrowing or broadening of the near-zero reflection region, and the reflection dip itself does not disappear. 

We have studied theoretically and also verified in experiment the effect of the total absorption by the dielectric metasurfaces supporting the transverse Kerker effect placed on a conducting substrate. The qualitative explanation of this effect is based on the Fabry-P\'erot resonances, and it shows that the absorption peak is due to a standing wave corresponding to the magnetic quadrupole resonance. More detailed analysis based on the metasurface multipole decomposition and the Airy formulas suggests a good agreement with both numerical simulations and experimental data, and it proves a  dominant role played  by the MQ resonance in the metasurface absorption. 

Thus, the dielectric metasurfaces governed by the transverse Kerker effect can provide both invisibility and full absorption, and they could be useful for a design of a variety of optical elements with enhanced properties allowing for more flexible manipulation of light with ultra-thin structures. Moreover, strong near-fields mediated by almost non-scattering regime pave the way to a plethora of  applications such as nonlinear harmonic generation, enhanced lasing and Raman scattering, and ultra-sensitive sensing. On the other hand, we believe that the perfect absorption delivered naturally by  low-absorbing dielectric particles could be of a significant relevance for the future device applications that require efficient light trapping and absorption, such as active integrated photonic circuits, selective thermal emitters, or microwave to infra-red signature controllers.

\section*{\label{ack}Acknowledgments}
HKS and AS contributed equally to this work. The experimental studies of the reflection and transmission coefficients in the microwave frequency range have been supported by the Russian Science Foundation (grant No. 17-19-01731). YSK acknowledges a support from the Strategic Fund of the Australian National University. VRT acknowledges a hospitality and financial support of the Jilin University. 

\bigskip

\bibliography{kerker_meta}

\end{document}